\documentclass[
aps,
prd,
eqsecnum,
notitlepage,
nofootinbib,
showpacs,
preprintnumbers,
superscriptaddress,
groupedaddress,
floatfix
]{revtex4-1}
\usepackage{amsmath,amssymb}
\usepackage{braket}
\usepackage{bibentry} 
\usepackage{color}
\usepackage{dsfont}
\usepackage{listings}
\usepackage{notes2bib} 
\usepackage{url}

\usepackage[
pdftex,
colorlinks=true,
linkcolor=blue,
filecolor=blue,
urlcolor=blue,
citecolor=blue,
plainpages=false,
bookmarksopen=true
]{hyperref} 

\DeclareFixedFont{\ttb}{T1}{txtt}{bx}{n}{9} 
\DeclareFixedFont{\ttm}{T1}{txtt}{m}{n}{9}  

\definecolor{deepblue}{rgb}{0,0,0.5}
\definecolor{deepred}{rgb}{0.6,0,0}
\definecolor{deepgreen}{rgb}{0,0.5,0}

\newcommand\pythonstyle{\lstset{
language=Python,
basicstyle=\ttm,
otherkeywords={self}, 
emph={__init__}, 
keywordstyle=\color{black},
commentstyle=\color{black},
emphstyle=\color{black}, 
stringstyle=\color{black},
frame=, 
showstringspaces=false
}}

\lstnewenvironment{python}[1][]
{
\pythonstyle
\lstset{#1}
}
{}

\newcommand\pythoninline[1]{{\pythonstyle\lstinline!#1!}}
\def\irrep{\pythoninline{irrep}}
\def\repobj{\pythoninline{rep_object}}
\def\mesiso{\pythoninline{meson_isospin_wrapper}}

\newcommand{\bnl}{
 Department of Physics,
 Brookhaven National Laboratory,
 Upton, NY, USA 11973
}

\begin{document}

\title{{\tt wickop}:
 Lattice Cubic Rotation Operator Generator}
\author{Aaron S. Meyer}
\email{ameyer@quark.phy.bnl.gov}
\affiliation{\bnl}
\date{\today}
\maketitle

\section{Introduction}
Many modern calculations in lattice QCD are computed with large operator bases,
 often including complicated interpolating operators with derivatives and
 multiple sets of quark bilinears.
These lattice QCD calculations access scattering matrix elements
 or characterize the properties of resonances
 such as the $\rho$ meson
 (for a recent review, see~Ref.~\cite{Briceno:2017max})
 or isolate exclusive scattering channels in the energy spectrum,
 such as in LQCD calculations of the muon hadronic vacuum polarization~\cite{
Erben:2017hvr,
Lehner:2018KEK,
Bruno:2019nzm}.
Isolating multiparticle scattering states in analysis is made easier by computing
 correlation functions with operators that couple strongly to the desired states.
However, creating the appropriate operators can often be fraught with difficulties,
 since translations and rotations are intimately linked.
Naive attempts to use tools for studying discrete groups are computationally
 costly due to a lattice volume factor in the group order of the translation group.
Getting around this difficulty requires less straightforward methods.

This document outlines the usage of a series of python 2.7 scripts designed to easily
 and efficiently create compound operators.
This is accomplished by computing tensor products of smaller building blocks
 that transform irreducibly under lattice cubic rotational and translational symmetry.
In particular, the code has the ability to handle representations of the cubic rotation group
 with any spatial momentum.
The key paradigm is to track the momentum separately from the rotations,
 fully utilizing the abelian structure of the translation subgroup.
This goes through the Wigner little group method to classify the subgroup of rotations
 that leave the momentum direction unchanged.
Using little groups circumvents the issue of a volume factor in the number
 of representation matrices that are needed,
 instead only requiring the 96 cubic rotation representation matrices.
Tensor products are computed with character tables and Clebsch-Gordan (CG) coefficients
 are saved for each decomposition in the tensor product.
The code also builds operators in a consistent basis,
 ensuring that operators that transform in the same representation have the same properties.

These scripts were designed to interface with the \pythoninline{wick} code~\cite{wick}.
This other code computes Wick contractions of the files generated by the \pythoninline{wickop}
 code, turning these operators into graphs of quark propagators.
The operators generated with \pythoninline{wickop} and contracted with \pythoninline{wick}
 were used in Ref.~\cite{Bruno:2019nzm} to compute two-pion correlation
 functions with various momentum combinations as well as a world-first computation
 of four-pion correlation functions in the isospin-1 channel.

\subsection{Installation}

Version 1.0 of the \pythoninline{wickop}
 code is publicly available through the github page in Ref.~\cite{wickop}.
It can be cloned from the repository by navigating to the desired directory
 and using the command:
\begin{python}
$ git clone https://github.com/asmeyer2012/wickop.git
\end{python}

There is no other actual installation needed to use this code.
To get started, simply import either the \pythoninline{reference_reps.py} or
 \pythoninline{class_scribe_isospin.py} files in the python interpreter
 or in a python script:
\begin{python}
>>> from reference_reps import *
\end{python}
The code in \pythoninline{class_scribe_isospin.py} is provided to demonstrate how the
 Clebsch-Gordan coefficients from tensor products can be written to
 human-readable plain-text files.
The examples in this script generate text files in a format designed
 for use with the code of Ref.~\cite{wick}.
It is likely that users will have individualized needs not fulfilled by
 the classes in \pythoninline{class_scribe_isospin.py},
 so most of the documentation is geared toward using the \repobj~class
 in the \pythoninline{class_irrep.py} file.

\section{User Entry Points}

\subsection{Representation Class Objects}
\subsubsection{Initialization}
Within the \pythoninline{class_irrep.py} source file, the \repobj~class is defined
 which is the raw representation object used to compute tensor products.
When creating an instance of the \repobj~class, two arguments must be supplied:
\begin{itemize}
\item \pythoninline{repGen}:
 A list of the representation matrices that generate the representation.
 Three representation elements are needed, $R_{z}$, $R_{x}$, and $P$, in that order.
 These objects must be 2-dimensional \pythoninline{np.ndarray} objects in a list.
\item \pythoninline{pList}:
 A list of the momenta that correspond to each row of the representation matrix.
 Each momentum should be a length-3 list of integer values.
 The same momenta can appear more than once,
  but each momentum within the same magnitude and class should appear the same number of times.
\end{itemize}

\subsubsection{{\tt list\_irreps()}}
The representation matrices that are supplied to the \repobj~class instance need not
 be irreducible.
They will be automatically decomposed when the \repobj~is initialized.
Once initialized,
 the irreps can be printed to the terminal with the \pythoninline{list_irreps()} command:
\begin{python}
>>> r0 = rep_object(repGen,pList)
>>> r0.list_irreps()
  0 000_a1+ [0, 0, 0]
\end{python}
The output displays three pieces of data on each line.
First is an index that corresponds to the order that the irreps are saved to internal memory.
The second is the internal name of the irrep,
 which includes the momentum class that the irrep belongs to and the distinct irrep within
 the little group of that momentum class.
The third is a representative choice of 3-vector momentum for this representation.

\subsubsection{{\tt compute\_tensor\_product()}}
To combine two \repobj s via a tensor product, all that is needed is to use
 the \pythoninline{compute_tensor_product()} function.
It takes a \repobj~as the only argument and assigns the irreps of that tensor product
 to a ``daughter'' representation.
For example, the zero-momentum tensor product of two $T^-_1$ representations
\begin{equation}
 T^-_1 \otimes T^-_1 = A^+_1 \oplus E^+ \oplus T^+_1 \oplus T^+_2 \,:
\end{equation}
\begin{python}
>>> r0 = rep_t1m_000.copy()
>>> r1 = rep_t1m_000.copy()
>>> r0.compute_tensor_product(r1)
>>> r0.list_irreps()
 daughter 0: 000_t1- [0, 0, 0]
   0 000_a1+ [0, 0, 0]
   1 000_ee+ [0, 0, 0]
   2 000_t1+ [0, 0, 0]
   3 000_t2+ [0, 0, 0]
\end{python}
In this example, ``daughter'' 0 is the daughter irrep obtained from the irrep decomposition
 when initializing the \pythoninline{rep_t1m_000} \repobj .
Had \pythoninline{r0} been a reducible representation when initialized, there would have
 been more than one daughter listed, and the irreps of each daughter printed in separate lists.

When two nonzero momentum irreps are combined,
 the resulting irreps can have many different momentum combinations.
Note that the tag in the second column denotes only the momentum class and little group
 irrep, not the actual momentum.
The reference momentum for each irrep with the correct magnitude is listed in the third column.
\begin{python}
>>> r0 = rep_a1x_100.copy()
>>> r1 = rep_a1x_100.copy()
>>> r0.compute_tensor_product(r1)
>>> r0.list_irreps()
 daughter 0: 100_a1x [0, 0, 1]
   0 100_a1x [0, 0, 2]
   1 000_a1+ [0, 0, 0]
   2 000_ee+ [0, 0, 0]
   3 000_t1- [0, 0, 0]
   4 110_a1x [1, 1, 0]
   5 110_a2x [1, 1, 0]
\end{python}

\pythoninline{compute_tensor_product()} may be called more than once on the same
 \repobj~to build more complicated operators.
For example, the product
\begin{equation}
 E^+ \otimes E^+ \otimes E^+
 \to (A_1^+ \oplus A_2^+ \oplus E^+) \otimes E^+
 \to (E^+) \oplus (E^+) \oplus (A_1^+ \oplus A_2^+ \oplus E^+)
\end{equation}
\begin{python}
>>> r0 = rep_eep_000.copy()
>>> r1 = rep_eep_000.copy()
>>> r0.compute_tensor_product(r1)
>>> r0.compute_tensor_product(r1)
>>> r0.list_irreps()
 daughter 0: 000_ee+ [0, 0, 0]
  daughter 0: 000_a1+ [0, 0, 0]
    0 000_ee+ [0, 0, 0]
  daughter 1: 000_a2+ [0, 0, 0]
    0 000_ee+ [0, 0, 0]
  daughter 2: 000_ee+ [0, 0, 0]
    0 000_a1+ [0, 0, 0]
    1 000_a2+ [0, 0, 0]
    2 000_ee+ [0, 0, 0]
\end{python}
However, note that once \pythoninline{compute_tensor_product()} has been called for a \repobj ,
 it can no longer be used as the argument of another \repobj 's
 call of \pythoninline{compute_tensor_product()}:
\begin{python}
>>> r0.list_irreps()
 daughter 0: 000_t1- [0, 0, 0]
   0 000_a1+ [0, 0, 0]
   1 000_ee+ [0, 0, 0]
   2 000_t1+ [0, 0, 0]
   3 000_t2+ [0, 0, 0]
>>> r0.compute_tensor_product(r0)
Traceback (most recent call last):
  File "<stdin>", line 1, in <module>
  File "~/wickop/class_irrep.py", line 117, in compute_tensor_product
    raise GroupTheoryError("cannot handle tensor product with daughters in
 second representation!")
defines.GroupTheoryError: cannot handle tensor product with daughters in
 second representation!
\end{python}

\subsubsection{{\tt compute\_full\_eigenvectors()}}
After a tensor product has been computed,
 it is useful to extract the Clebsch-Gordan
 coefficients giving the linear combinations
 that transform irreducibly under the symmetry group.
These are extracted from the top-level \repobj~by calling
 \pythoninline{compute_full_eigenvectors()} with no arguments:
\begin{python}
>>> r0 = rep_eep_000.copy()
>>> r1 = rep_t1p_000.copy()
>>> r0.compute_tensor_product(r1)
>>> r0.list_irreps()
 daughter 0: 000_ee+ [0, 0, 0]
   0 000_t1+ [0, 0, 0]
   1 000_t2+ [0, 0, 0]
>>> evec,irrep,refmom = r0.compute_full_eigenvectors()
>>> irrep
['000_t1+', '000_t2+']
>>> refmom
[[0, 0, 0], [0, 0, 0]]
>>> evec[0]
array([[ 0.5      +0.j,  0.       +0.j,  0.       +0.j],
       [ 0.       +0.j,  0.5      +0.j,  0.       +0.j],
       [ 0.       +0.j,  0.       +0.j, -1.       +0.j],
       [-0.8660254+0.j,  0.       +0.j,  0.       +0.j],
       [ 0.       +0.j,  0.8660254+0.j,  0.       +0.j],
       [ 0.       +0.j,  0.       +0.j,  0.       +0.j]])
\end{python}

The return values of \pythoninline{compute_full_eigenvectors()}
 include the vectors of CG coefficients as well as the names and reference momenta
 of the irreps corresponding to each set of vectors.
The order of irreps in the output is the same as what is printed by
 \pythoninline{list_irreps()}.
In the example above, \pythoninline{evec[0]} contains the CG coefficients
 for the $T^+_1$ irrep obtained from the tensor product $E^+ \otimes T^+_1$.
There are three vectors which are arranged in columns.
The vectors are computed and ordered according to the code's internal conventions for the basis.
These entries correspond to the CG coefficients
\begin{equation}
 \pythoninline{evec[0]} = \left( \begin{array}{ccc}
 \braket{T^+_1[0] \,|\, E^+[0] T^+_1[0]} &
 \braket{T^+_1[1] \,|\, E^+[0] T^+_1[0]} &
 \braket{T^+_1[2] \,|\, E^+[0] T^+_1[0]} \\
 \braket{T^+_1[0] \,|\, E^+[0] T^+_1[1]} &
 \braket{T^+_1[1] \,|\, E^+[0] T^+_1[1]} &
 \braket{T^+_1[2] \,|\, E^+[0] T^+_1[1]} \\
 \braket{T^+_1[0] \,|\, E^+[0] T^+_1[2]} &
 \braket{T^+_1[1] \,|\, E^+[0] T^+_1[2]} &
 \braket{T^+_1[2] \,|\, E^+[0] T^+_1[2]} \\
 \braket{T^+_1[0] \,|\, E^+[1] T^+_1[0]} &
 \braket{T^+_1[1] \,|\, E^+[1] T^+_1[0]} &
 \braket{T^+_1[2] \,|\, E^+[1] T^+_1[0]} \\
 \braket{T^+_1[0] \,|\, E^+[1] T^+_1[1]} &
 \braket{T^+_1[1] \,|\, E^+[1] T^+_1[1]} &
 \braket{T^+_1[2] \,|\, E^+[1] T^+_1[1]} \\
 \braket{T^+_1[0] \,|\, E^+[1] T^+_1[2]} &
 \braket{T^+_1[1] \,|\, E^+[1] T^+_1[2]} &
 \braket{T^+_1[2] \,|\, E^+[1] T^+_1[2]} \\
 \end{array} \right) \,.
\end{equation}
Additional tensor products will result in products of CG coefficients
 for every tensor product in the chain, e.g.
\begin{equation}
 [([E^+ \otimes T^+_1] \to T^+_1) \otimes E^+] \to T^+_1 \quad:\quad
 \braket{T^+_1[k] \,|\, E^+[i] T^+_1[j]}
 \braket{T^+_1[m] \,|\, T^+_1[k] E^+[\ell] } \,.
\end{equation}

\subsection{Meson Isospin Wrapper}
Within the \pythoninline{class_scribe_isospin.py} source file
 exists a class for composing isospin-1 meson operators into multiparticle operators.
The class is named \mesiso .
This pairs the code for computing CG coefficients with an implementation of
 a class for mapping elements of a representation into human-readable text.
Some example code using \mesiso~is available at the end of
 \pythoninline{class_scribe_isospin.py}.
This class is meant to be an example of how to pair the
 \repobj~class with output text and is designed for use with
 the {\ttm wick} codebase~\cite{wick}.

Initialization of the \mesiso~class just takes
 a keyword dictionary as input, e.g.
\begin{python}
>>> for key in sorted(kwargs.keys()):
...  print '%20s %30s' %(key, kwargs[key])
... 
      deriv_operator                           None
               gamma           ['GammaI', 'GammaI']
 hermitian_conjugate                          False
  reference_momentum         [[0, 0, 0], [0, 0, 0]]
>>> mi0 = meson_isospin_wrapper(**kwargs)
-starting group theory computation
-starting tensor products
-- tensor product 0
-done with group theory computation
-time: 0.08
\end{python}
This example creates operators which are the tensor product of two zero-momentum
 isospin-1 vector mesons, $\bar{\psi} \gamma_i \psi$.
The keyword inputs can be understood as follows:
\begin{itemize}
\item \pythoninline{deriv_operator}:
Keyword for specification of derivative operators with different symmetry transformations.
This is handled by mapping the name of the derivative operator to one
 of the irreps given in \pythoninline{defines.py} by means of the dictionary
 \pythoninline{switch_deriv_repname}.
If this is not \pythoninline{None}, then must be a list with length matching
 the other keyword arguments.
Leaving \pythoninline{deriv_operator} as \pythoninline{None} ignores this modifier
 and uses only the momentum and gamma structure.
\item \pythoninline{gamma}: 
List of gamma structures to use for each meson operator building block.
These are mapped to representations via another dictionary,
\begin{python}
switch_gamma_rep = {
 'Identity':     rep_a1p_000,
 'GammaT':       rep_a1p_000,
 'GammaI':       rep_t1m_000,
 'GammaIGammaT': rep_t1m_000,
 'GammaIGammaJ': rep_t1p_000,
 'GammaIGamma5': rep_t1p_000,
 'GammaTGamma5': rep_a1m_000,
 'Gamma5':       rep_a1m_000 }
\end{python}
\item \pythoninline{hermitian_conjugate}:
Flag that changes the behavior of the scribe to write out the Hermitian conjugate
 rather than the unmodified operator.
\item \pythoninline{reference_momentum}:
List of reference momentum to use for each meson operator building block.
These are length-3 lists of integers corresponding to the momenta.
The number of momenta here must match the number of gamma structures assigned
 to the keyword \pythoninline{gamma}.
\end{itemize}
The initialization of the \mesiso~class
 combines the representation matrices for the momentum and Dirac gamma structure
 and creates the required generators of the symmetry group.
It also fills out a list of momenta.
Both of these are used as inputs to the \repobj~class object that does the
 group theory computation.

One last command exists to write the operators to files.
This is the \pythoninline{generate()} command:
\begin{python}
>>> mi0.generate()
- generate_isospin_term time: 0.03
- generate_isospin_term time: 0.05
- generate_isospin_term time: 0.05
- generate_isospin_term time: 0.06
- generate_isospin_term time: 0.07
- generate_isospin_term time: 0.08
- generate_isospin_term time: 0.09
- generate_isospin_term time: 0.09
- generate_isospin_term time: 0.1
- generate_isospin_term time: 0.11
- generate_isospin_term time: 0.11
- generate_isospin_term time: 0.11
- generate_isospin_term time: 0.12
- generate_isospin_term time: 0.12
- generate_isospin_term time: 0.13
- generate_isospin_term time: 0.13
- generate_isospin_term time: 0.14
- generate_isospin_term time: 0.15
- generate_isospin_term time: 0.16
- file write time: 0.4
\end{python}
This command will create a directory structure within the \pythoninline{wickop} directory
 that separates the operators according to their spin structure, center-of-mass momentum,
 and isospin components.
The file names also contain the representation name, the component of the representation,
 and a (meaningless) unique index.
The example above generates the file
\begin{python}
output_baseop00/i1c0/pcom000/gigip000p000/t1p_000.00002.00.op
\end{python}
The top directory is a generic directory name where all of the operators are written.
Second is the isospin (1) and the isospin $I_z$ component (0),
 followed by the center-of-mass momentum ($\vec{p}=(0,0,0)$).
Since this operator was made up of two bilinears with $\gamma_i$ Dirac structure
 and $\vec{p}=(0,0,0)$, the next directory indicates that construction.
Finally, this is the $T^+_1$ representation of the $\vec{p}=(0,0,0)$ momentum class.
The five-digit index is the identifying index, and \pythoninline{00}
 refers to the 0-component of the representation,
 which in the code's basis refers to the operator along the $x$-axis.
Many such files are written when calling the \pythoninline{generate()} command.
In this particular example, 81 files are generated.

The syntax of the output file is as follows:
\begin{python}
$ cat output_baseop00/i1c0/pcom000/gigip000p000/t1p_000.00002.00.op

# 000_t1+ p=0,0,0 LG_index=0

FACTOR -0.5
 UBAR t local
 GAMMA 1
 D t local
 DBAR t local
 GAMMA 2
 U t local

FACTOR 0.5000000000000001
 UBAR t local
 GAMMA 2
 D t local
 DBAR t local
 GAMMA 1
 U t local

# 000_t1+ p=0,0,0 LG_index=0

FACTOR 0.5
 DBAR t local
 GAMMA 1
 U t local
 UBAR t local
 GAMMA 2
 D t local

FACTOR -0.5000000000000001
 DBAR t local
 GAMMA 2
 U t local
 UBAR t local
 GAMMA 1
 D t local
\end{python}
The file contains the ``00'' ($x$) component of the $T^+_1$ representation,
 which is the operator combination
\begin{equation}
  \frac{1}{2}\Big[-(\bar{u}\gamma_y d)(\bar{d}\gamma_z u)
                  +(\bar{u}\gamma_z d)(\bar{d}\gamma_y u)\Big]
+ \frac{1}{2}\Big[+(\bar{d}\gamma_y u)(\bar{u}\gamma_z d)
                  -(\bar{d}\gamma_z u)(\bar{u}\gamma_y d)\Big] \,.
\end{equation}
The keyword ``local'' is a modifier that can be altered based on the desired application.

\section{Evaluation of Tensor Products}

Evaluation of the tensor products and the irrep decomposition applies many of the usual methods
 for decomposing representations of discrete groups.
Projection matrices are constructed from the character tables of the octahedral group 
 (or a corresponding little group) and the full set of representation matrices in that group.
These projections are used to deduce the Clebsch-Gordan coefficients for the products.

The main complication that arises from these decompositions is the handling of
 operator momentum combinations.
The full set of group elements includes all powers of single-site translations,
 which results in a lattice-volume factor increase in the order of the group.
Even though the translation subgroup is abelian,
 rotations do not commute with translations and so nonzero momentum representations
 will have nontrivial properties under rotations.

Fortunately, the abelian translation subgroup also permits shortcuts
 in how the translations are handled.
Instead, momenta may be tracked as a separate index that also transforms under
 the cubic rotational symmetry,
\begin{equation}
 \hat{R} \ket{e_i(\vec{p})} = R_{ij} \ket{e_j(R^{-1}\vec{p})} \,,
\end{equation}
 where $e_i$ is a basis vector in the rotational little group for momentum $\vec{p}$.
Rather than computing representation matrices for the full set of group elements,
 the subset corresponding to only transformations within the little group must be considered.
The remaining basis vectors are reconstructed from a coset of representation elements
 from the rotation octahedral group.
When constructed in this way, all basis vectors are automatically eigenstates
 of the translations,
\begin{equation}
 \hat{T}^n_j \ket{e_i(\vec{p})} = e^{-i n p_j} \ket{e_i(\vec{p})} \,,
\end{equation}
 and no translation representation matrices must ever be constructed.

The code is designed to take two irreducible representations as input
 and produce the Clebsch-Gordan coefficients relating the product
 states to every irrep in the tensor product decomposition.
The starting point is the set of representation matrices that generate the
 rotation subgroup and the momentum of the states corresponding to every row
 in the representation matrices.
These are combined by a matrix tensor product and then decomposed.
The engine determines a set of basis vectors that transform irreducibly under
 the full symmetry group.
The basis vectors may be applied to the representation matrices to form
 a new set of generators with rank equal to the dimension of the irrep.
These are the generators of the new irrep.
Later tensor products are computed from the new generators,
 and any operators built up from the full tensor product series are iteratively
 constructed from the basis vectors at each level of the decomposition.

This section gives the mathematical formulae that are applied to carry out the irrep decomposition.
Each subsection details a separate step in the decomposition process.
The results of each decomposition are saved in an internal \irrep~class
 that contains the information about the irrep, momentum, and projecting vectors.
Subsequent tensor products are applied to the lowest-level \irrep~classes
 in the chain of tensor products, using the minimal necessary information.
All of this data is encapsulated in a \repobj~class,
 and the user never needs to interface with the underlying \irrep~classes.

\subsection{Representation Object Internals}

When a \repobj~class is initialized, a list of the representation matrices
 \pythoninline{repGen}$\; = [R_{z}, R_{x}, P]$, is required.
These are the rotations about the $z$- and $x$-axes as well as a parity transformation.
The second input is a list of momenta for this representation,
 which is the momentum of each row and column of the representation matrix.
To make this explicit, the representation matrix $R$ is written in terms
 of basis vectors $\ket{v_i}$ as
\begin{equation}
 R = \sum_{ij} \ket{v_i} \bra{v_j} R_{ij} \,,
\end{equation}
 where the basis vectors have momenta defined in \pythoninline{pList}:
\begin{equation}
 \pythoninline{pList}[i] = \vec{p}\,^{(i)}:
 \quad \hat{T}_j \ket{v_i} = e^{-i p_j^{(i)}} \ket{v_i} \,.
\end{equation}
The momentum list will be designated ${\cal P}$ for short,
 with \pythoninline{pList}$[i] \equiv {\cal P}_i$.
The representation matrices $R_{z}$, $R_{x}$, and $P$ is the minimal set necessary
 to build the full set of representation matrices for the order-96 
 double cover of the octahedral group.
No translations are ever considered, as all of the pertinent information is contained
 in the momenta.

After initializing the \repobj~class,
 an initial decomposition is computed to reduce the input representation matrices to irreps.
These are stored internally as \irrep~classes, which are operated on in later steps.

\subsection{Matrix Tensor of Generators and Momentum and Presorting}

The tensor product decomposition of two irreducible representations $\rho_1$ and $\rho_2$
 starts with the tensor product of its representation matrices for the generators $R$,
\begin{equation}
 \rho_1(R) \otimes \rho_2(R) = \rho_{1\otimes2}(R) \,.
\end{equation}
This is composed of the product of states $\ket{v^{(1)}_i}$ from $\rho_1$ and
 $\ket{v^{(2)}_i}$ from $\rho_2$,
 which have momenta ${\cal P}^{1}_i$ and ${\cal P}^{2}_i$, respectively.
The product state $\ket{v^{(1)}_i} \ket{v^{(2)}_j}$ then has momentum
 ${\cal P}^{1\otimes 2}_{i,j} = {\cal P}^1_i +{\cal P}^2_j$.
The representation matrices are then sorted by momenta ${\cal P}^{1\otimes 2}_{i,j}$,
 first sorting by the magnitude $|{\cal P}^{1\otimes 2}_{i,j}|=|\vec{p}|$
 and then canonically sorting the momentum directions within the same momentum class.
The result is a permutation matrix $\Gamma$, which is applied via a similarity transform,
\begin{equation}
 \Gamma^T \rho_{1\otimes2}(R) \Gamma
 = \bigoplus_{(|\vec{p}|,\hat{p})} \rho_{(|\vec{p}|,\hat{p}) \in {\cal P}^{1\otimes2}}(R) \,.
\end{equation}

The resulting matrix $\rho_{(|\vec{p}\,|,\hat{p}) \in {\cal P}^{1\otimes2}}(R)$
 is partially block-diagonalized into generally reducible but independent subblocks.
The advantage of this partial diagonalization is speed.
Each subblock may be independently decomposed by character projection.
In addition, different momentum classes are separated from each other,
 so the rotational little group is known {\it a priori}.

Suppose a \repobj~instantiation is created for representations $\rho_1$ and $\rho_2$,
 named \pythoninline{r1} and \pythoninline{r2} for concreteness.
When \pythoninline{r1.compute_tensor_product(r2)} is called,
 a new \repobj~is constructed for every pair of \irrep~classes,
 one taken from \pythoninline{r1.irreps} and the other from \pythoninline{r2.irreps}.
The new \repobj s are initialized with the matrix tensor product of the
 \pythoninline{repGen} attributes of each and the sums of momenta in the \pythoninline{pList}
 attribute.
The new \repobj s are saved in the attribute \pythoninline{rep_object.daughters}
 of \pythoninline{r1}.
The permutation matrix $\Gamma$ is stored internally as the attribute
 \pythoninline{r1.perm} in the corresponding \repobj.

If the \repobj~\pythoninline{r1} was previously called for a tensor product,
 then \pythoninline{rep_object.compute_tensor_product(r2)} is applied iteratively to all
 \repobj s saved in the attribute \pythoninline{r1.daughters} instead of to the
 \irrep s in \pythoninline{r1}.

\subsection{Group Theoretic Projection Formula}

This step builds projection matrices to decompose the subblocks of
 $\rho_{(|\vec{p}|,\hat{p}) \in 1\otimes2}(R)$.
To isolate irreps, the group theoretic projection formula is applied,
\begin{equation}
 P_{\sigma} = \frac{\dim{\sigma}}{|G|} \sum_{g\in G} \chi_{\sigma}^{\ast}(g) \rho(g) \,.
 \label{eq:irrepprojection}
\end{equation}
Here, $|G|$ is the order of the group, 
 $\rho(g)$ is a generally reducible representation matrix for group element $g$,
 $\chi_\sigma(g)$ is the character of irrep $\sigma$,
 and $\dim{\sigma}$ is the dimension of the irrep.
The projection formula is applied only to the subblock with definite reference momentum
 $\vec{p}=\vec{p}_{ref}$ and only for the little group associated to that $\vec{p}_{ref}$.
The specific choices of reference momenta are described in Appendix~\ref{sec:refmom}.
This is only a subblock of the subblock represented by
 $\rho_{(|\vec{p}|,\hat{p}) \in 1\otimes2}(R)$.
In the case of nonzero momentum,
 $|G|$ is then the little group associated to reference momentum $\vec{p}_{ref}$
 and $\sigma$ is an irrep of the little group rather than the full octahedral group.

Each representation $\sigma$ is computed in a loop, and all of the following steps are applied.
If the irrep $\sigma$ is not represented by the decomposition, then the resulting
 projection matrix is a zero matrix and the following steps are skipped.
The projection matrix is not saved internally,
 since it is redundant with the eigenvectors that are computed in later steps.

\subsection{Eigenvector Solutions of Basis Choice}

The basis must be chosen consistently for each irrep.
Some basis elements are fixed by computing eigenvectors of representation generators
 that have been projected to irreps using Eq.~(\ref{eq:irrepprojection}).
The rest of the basis elements are computed from the initial set of basis vectors later on.
While this guarantees a consistent basis, the downside is that
 the choice of basis must be explicitly specified for each irrep.

The eigenvector equation to fix basis vectors is
\begin{equation}
 \rho_{(|\vec{p}|,\hat{p}) \in 1\otimes2}(R) P_{\sigma} \bar{V}_n = \lambda_n \bar{V}_n \,,
 \label{eq:irrepeigenvector}
\end{equation}
 which uses the irrep projection matrix from Eq.~(\ref{eq:irrepprojection}).
The representation matrix $R$ and eigenvalue $\lambda_n$ are chosen
 differently for each $\sigma$, and these choices are given in Appendix~\ref{sec:lgbasis}.
One conventional choice is, for example,
 $R=R_z$ and $\lambda_n = 1$ for the $T^-_1$ representation,
 which singles out the operator that is invariant under a $z$-axis rotation.

In general, there could be more than one solution to the eigenvector equation
 if the same irrep appears more than once in a tensor product decomposition.
In such cases, there is a degeneracy that can be utilized to simplify the operators.
If $\bar{V}^{(m)}_n$ is a set of $M$ length-$N$ column vectors,
 then the operators are fixed by contracting with a length-$N$ vector $Q^{m'm}$,
\begin{equation}
 \bar{V}^{(m')}_n Q^{m'm} = V^{(m)}_n \,,
\end{equation}
 where $m$ is an index in the $M$ independent vector subspaces.
These subspaces cannot be mixed by application of representation elements,
 a property that is utilized when computing other basis vectors.
The vectors $Q^{m'm}$ are chosen to reduce the number of nonzero entries in the $V^{m}$s.
For an $M$-fold irrep degeneracy with $m \in \{0,\dots,M-1\}$, it is possible to zero out $M-m-1$
 linearly-independent entries of the vector $V^{m}$ without zeroing out the entire vector.
This is carried out with Gauss-Jordan elimination over nonzero entries of $V^{m}$.

\subsection{Remaining Vector Phases of Basis Choice}

The remaining basis vectors are determined from the initial basis vector
 fixed by solving the eigenvector equation and simplification in the previous section.
The missing basis vectors are obtained by applying representation matrices
 to the eigenvectors and requiring certain phase conventions under those transformations.
This is done first for the little group,
 where all of the basis vectors have the same momentum,
 and then again to get all momentum combinations that are connected by rotations.

The eigenvectors determined in the previous section all transform irreducibly,
 so all that is needed to get the remaining basis vectors is to apply the
 representation matrices to them.
For the basis vectors with $\vec{p} = \vec{p}_{ref}$,
 the remaining vectors can be obtained from the relation
\begin{equation}
 \sum_{n'} \rho_{nn'}(R) V_{n',i} = \sum_{j} V_{n,j} a_{ji}
 \label{eq:basisfromrotation}
\end{equation}
 where $R$ is a group element belonging to the little group.
The little group for each momentum class is known,
 so $a_{ji}$ can be conventionally chosen {\it a priori}.
This fixes the transformations within the little group.
For the basis vectors outside of the little group, i.e. with $\vec{p} \neq \vec{p}_{ref}$,
 Eq.~(\ref{eq:basisfromrotation}) is applied again to every basis vector in
 the little group but with the requirement that $a_{ji}$ is $1$
 for only one entry and $0$ otherwise.
The representation element $R$ for these transformations is one of the coset elements,
 given in Appendix~\ref{sec:cosets}.

At this point, the irrep decomposition is complete.
The basis vectors $V^{m}_{n,i}$ are then the CG coefficients for the decomposition
 of the tensor product into an irrep:
\begin{equation}
 V^{(m)}_{n,i} =
 \sum_{n_1=0}^{(\dim{\rho_1}-1)} \sum_{n_2=0}^{(\dim{\rho_2}-1)}
 \braket{\sigma^{(m)},i|\rho_1,n_1;\rho_2,n_2}
 \delta_{n,(n_1\cdot \dim{\rho_2} +n_2)} \,.
\end{equation}
The replica copies of the irrep $\sigma$, denoted by the superscript $m$,
 each have their own set of orthogonal CG coefficients,
\begin{equation}
 \sum_{n_1,n_2}
 \braket{\sigma^{(m)},i|\rho_1,n_1;\rho_2,n_2}
 \braket{\rho_1,n_1;\rho_2,n_2|\sigma^{(m')},j} = \delta^{mm'}\delta_{ij} \,.
\end{equation}

The basis vectors are saved to \irrep~classes under the \pythoninline{irrep.evec}
 attribute of each class object.
The \irrep~classes are held in the \pythoninline{repobj.irreps} attribute.
The attribute is saved as \pythoninline{irrep.evec[n][i]} $=V^{(m)}_{n,i}$
 for fixed $m$.
When the irrep multiplicity is larger than one,
 different choices of $m$ are contained in different \irrep~class instances.

\subsection{Building the Irrep Representation Matrices}

If more than one tensor product is computed,
 the code will use the \irrep~class objects of the daughter representations
 rather than the full representation matrix.
This reduces the computation workload, since the irrep matrices are considerably smaller.
The representation matrices can be deduced from the full representation matrices
 by projecting down to the subspace of the basis vectors.
The resulting matrix is of dimension $(\dim{\sigma}\times\dim{\sigma})$,
\begin{equation}
 [\Gamma V^{(m)}]^\dagger \rho_{1\otimes2}(R) [\Gamma V^{(m)}] = \sigma^{(m)}(R) \,.
\end{equation}
This is the representation matrix that is used in the tensor products.
The representation matrices for the same irrep are always identical,
 i.e. $\sigma^{(m)}(R)=\sigma^{(m')}(R)$ for $m\neq m'$.

The full representation matrix can also be projected only to the invariant subspace.
This is simply
\begin{equation}
 \rho_{1\otimes2}(R) [\Gamma V^{(m)}] [\Gamma V^{(m)}]^\dagger = \rho_{\sigma^{(m)}}(R) \,,
\end{equation}
 since the outer product of the basis vectors just gives the projection matrix itself,
\begin{equation}
 P_{\sigma^{(m)}} = [\Gamma V^{(m)}] [\Gamma V^{(m)}]^\dagger \,.
\end{equation}

Even when several tensor products are chained together,
 it is not necessary to save the basis vectors including the indices of all of
 subrepresentations.
The basis vectors in the full product space are reconstructed on the fly from the
 basis vectors for each tensor product in the chain.
For example, if the tensor product chain
\begin{equation}
 [([\rho_1 \otimes \rho_2] \to \sigma) \otimes \rho_3] \to \sigma'
\end{equation}
 is computed, with basis vectors $[\Gamma V]_1$ for $[\rho_1 \otimes \rho_2] \to \sigma$
 and $[\Gamma V]_2$ for $[\sigma \otimes \rho_3] \to \sigma'$,
 then the full basis vectors for 
 $[\rho_1 \otimes \rho_2 \otimes \rho_3] \to \sigma'$ are
\begin{equation}
 \Omega = [\Gamma V]_1 [\Gamma V]_2 \,.
\end{equation}
The matrix $\Omega$ can be used to build the projection to $\sigma'$,
\begin{equation}
 \rho_{1\otimes2\otimes3}(R)
 \Omega \Omega^\dagger
 = \rho_{1\otimes2\otimes3\to\sigma'}(R)
\end{equation}
The full set of vectors $\Omega$ contains redundant information,
 so only the CG coefficient vectors for each step in the chain are saved.

\section{Conclusions}

The production of operators with definite translational and rotational symmetry
 is a nontrivial task.
The relationship between translations and rotations that is
 designated by a semidirect product leads to a rich interplay between the two symmetries.
Combinations of multiple operators with momentum can lead to other operators that have
 nontrivial transformations under rotations,
 or momentum and rotations could conspire to give simpler transformations under the symmetries.

All of the rich structure that is present in this group also makes it difficult to 
 combine operators quickly via tensor product relations.
Representations depend on the total momentum of the operator as well as any
 operator structure that transforms under rotations.
These representations may be naively computed by writing all of the representation matrices
 and using the tools for studying discrete groups.
This builds a volume factor into the computation, making the tensor product decompositions slow.

The \pythoninline{wickop} codebase exploits the abelian nature of the translations
 to shortcut the volume dependence in the number of representation elements.
Instead, representations are computed using only the 96 cubic rotation group
 representation elements, tracking momentum as an independent quantity.
This greatly speeds up the computation time and allows for easy generation
 and combination of operators with known symmetry properties.
With these scripts, it is possible to easily generate operators with definite
 momentum, spin, and isospin and write these operators in a human-readable format.
When combined with the \pythoninline{wick} codebase,
 one can easily compute contractions of operators and build up complicated graphs
 involving many quark lines, easing the analysis effort and offering a faster,
 more robust workflow.

\subsection{Acknowledgements}

I would like to thank 
 Mattia Bruno,
 Zechariah Gelzer,
 Ciaran Hughes,
 Andreas Kronfeld,
 Christoph Lehner,
 Yin Lin,
 and my colleagues in the Fermilab, RBC, and UKQCD collaborations for interesting discussions.
This work was supported by Brookhaven Science Associates, LLC under
 Contract No. DE-SC0012704 with the U.S. Department of Energy.

\appendix

\section{Symmetry Group}

The continuum, infinite-volume, nonrelativistic symmetry group of particles
 with angular momentum is
\begin{equation}
 G_{C} = \mathds{R}^3 \rtimes (\text{SU}(2) \times \mathds{Z}_2) \,.
 \label{eq:continuumgroup}
\end{equation}
This set of symmetry transformations contains translations, rotations, and parity operations.
For a discrete lattice timeslice with infinite volume, the symmetry group is
\begin{equation}
 G_{L_\infty} = \mathds{Z}^3 \rtimes \widetilde{\text{W}}_3 \,.
 \label{eq:latticegroupinfinite}
\end{equation}
The discrete group $\widetilde{\text{W}}_3$ is the double cover of the octahedral
 group and includes the parity transformation as well as $\frac{\pi}{2}$
 rotations about the lattice axes.
On a finite lattice, the continuous translational symmetry $\mathds{R}^3$
 is broken down into the discrete translation symmetry $\mathds{Z}^3$.
If instead a periodic, finite volume lattice of size $L$ is considered,
 the translation symmetry group factor is replaced with cyclic translations,
\begin{equation}
 G_{L} = (\mathds{Z}_L)^3 \rtimes \widetilde{\text{W}}_3 \,.
 \label{eq:latticegroupfinite}
\end{equation}
The only difference between representations of the groups in
 Eqs.~(\ref{eq:latticegroupinfinite})~and~(\ref{eq:latticegroupfinite})
 comes from momenta near the Brillouin zone,
 which are not generally probed by lattice calculations.
If the momenta are not that large,
 then the representation theory of the groups is the same.
The \pythoninline{wickop} code assumes the symmetry group in Eq.~(\ref{eq:latticegroupinfinite}).

When considering representations that transform in the discrete lattice subgroup,
 it is preferable to work directly with the symmetry group in question.
From the generators of the discrete group, one may wish to use the standard methods of characters
 and projections to do tensor product representation decomposition.
However, dealing with the infinite (or even finite) translation subgroup can be messy,
 and significantly adds to the computing requirements since it increases the 
 order of the group by a factor of $L^3$.
Fortunately, the translation subgroup is abelian.
Representations of the full translation-rotation-parity group can be lifted
 from the little group rotation representation for a single, fixed reference momentum.

\begin{table}
\centering
 \begin{tabular}{c|c|c|c|c|c}
  $\vec{p}$ class & \#$\vec{p}$ & $\Gamma$ & $\gamma\in\Gamma$
  & dim$[\gamma]$ & dim$[\gamma \uparrow G]$ \\
\hline
  $(0,0,0)$ & 1 & $\widetilde{\text{W}}_3$ &
  $\begin{array}{c}
  A_1^\pm, A_2^\pm \\
  E^\pm \\
  T_1^\pm, T_2^\pm \\
 \hline
  G_1^\pm, G_2^\pm \\
  H^\pm \end{array}$ & 
  $\begin{array}{c} 1\\ 2\\ 3\\ \hline 2\\ 4 \end{array}$ &
  $\begin{array}{c} 1\\ 2\\ 3\\ \hline 2\\ 4 \end{array}$ \\
\hline
  $(0,0,p)$ & 6 & $\widetilde{\text{D}}_4 \simeq \mathds{Q}_{16}$ &
  $\begin{array}{c}
  A_1, A_2, B_1, B_2 \\
  E_2 \\
 \hline
  G_1, G_2 \end{array}$ & 
  $\begin{array}{c} 1\\ 2\\ \hline 2 \end{array}$ &
  $\begin{array}{c} 6\\12\\ \hline12 \end{array}$ \\
\hline
  $(p,p,0)$ & 12 & $\widetilde{(\mathds{Z}_2\times\mathds{Z}_2)} \simeq \text{D}_4$ &
  $\begin{array}{c}
  A_1, A_2, B_1, B_2 \\
 \hline
  G_1 \end{array}$ & 
  $\begin{array}{c}  1\\ \hline  2 \end{array}$ &
  $\begin{array}{c} 12\\ \hline 24 \end{array}$ \\
\hline
  $(p,p,p)$ & 8 & $\widetilde{\text{D}}_3 \simeq \text{Dic}_{12}$ &
  $\begin{array}{c}
  A_1, A_2 \\
  B \\
 \hline
  L_1, L_2 \\ 
  G_1 \end{array}$ & 
  $\begin{array}{c} 1\\  2\\ \hline 1\\  2 \end{array}$ &
  $\begin{array}{c} 8\\ 16\\ \hline 8\\ 16 \end{array}$ \\
\hline
  $(p,q,0)$ & 24 & $\widetilde{\mathds{Z}}_2 \simeq \mathds{Z}_{4}$ &
  $\begin{array}{c}
  A_1, A_2 \\
 \hline
  L_1, L_2 \end{array}$ & 
  $\begin{array}{c} 1\\ \hline 1 \end{array}$ &
  $\begin{array}{c}24\\ \hline24 \end{array}$ \\
\hline
  $(p,p,q)$ & 24 & $\widetilde{\mathds{Z}}_2 \simeq \mathds{Z}_{4}$ &
  $\begin{array}{c}
  A_1, A_2 \\
 \hline
  L_1, L_2 \end{array}$ & 
  $\begin{array}{c} 1\\ \hline 1 \end{array}$ &
  $\begin{array}{c}24\\ \hline24 \end{array}$ \\
\hline
  $(p,q,r)$ & 48 & $\widetilde{\mathds{1}} \simeq \mathds{Z}_{2}$ &
  $\begin{array}{c}
  A_1 \\
 \hline
  L_1 \end{array}$ & 
  $\begin{array}{c} 1\\ \hline 1 \end{array}$ &
  $\begin{array}{c}48\\ \hline48 \end{array}$ \\
 \end{tabular}
 \caption{
 List of the full set of momentum classes and irreducible representations within
  each momentum class.
 The first column is the momentum class, and the second column is the number
  of momenta in that class that can be reached by symmetry transformations.
 The third column is the little group $\Gamma$ associated with that momentum class.
 For each little group, both the group that has a double cover and the group isomorphic
  to it are shown.
 These are:
 $\widetilde{\text{W}}_3$, the octahedral rotation group with parity;
 $D_{n}$, the dihedral group of order $2n$;
 $\mathds{Q}_{16}$, the quaternion group of order 16;
 $\text{Dic}_{12}$, the dicyclic group of order 12;
 and $\mathds{Z}_n$, the cyclic group of order $n$.
 In the fourth column, the names of the little group irreps $\gamma$ are given,
  separated into bosonic and fermionic representations by a horizontal line,
  and the fifth column gives the dimensions of the irreps in that little group representation.
 The irrep dimensions of the representations in the full translation-rotation-parity group $G$
  are the product of the $\#\vec{p}$ column with the $\text{dim}[\gamma]$ column,
  given in the last column $\text{dim}[\gamma\uparrow G]$.
 }
\end{table}

\section{Conventions}

\subsection{Reference Momenta}
\label{sec:refmom}

The choice of reference momentum is a matter of convention and does not affect the
 resulting operators apart from a choice of basis.
However, consistency is important to ensure that the basis generated by the code
 is the same for construction of operators in the same irrep.
This is because the choice of coset elements used to generate the full set of
 representation basis vectors assumes a specific choice of reference momentum.
For this reason, the conventions for choosing reference momenta are outlined below.

There are seven possible classes that the momenta can fall into.
These are labeled by the 3-vectors $(0,0,0)$, $(0,0,p)$, $(p,p,0)$, $(p,p,p)$,
 $(p,q,0)$, $(p,p,q)$, and $(p,q,r)$, where $p$, $q$, and $r$ are all distinct.
The reference momentum for the zero momentum case is trivial since there is only one momentum,
 $\vec{p}_{ref} = 0$.
The other choices will be explained below.

When choosing a reference momentum during the computation,
 some conventional choices are considered.
First, the reference momenta always take all nonzero momentum directions to be positive.
In addition, the reference momenta for each momentum class are chosen to
 favor the $z$-axis direction whenever possible.
This is the case for all nonzero momenta except for $(p,p,p)$ and $(p,q,r)$.
For the former, there is only one choice with all directions positive.
The latter is ordered so that $p > q > r$.

The cases $(p,q,0)$ and $(p,p,q)$ can still be ambiguous.
The first is chosen to order the momenta, again choosing $p > q > 0$.
The second fixes the momenta so that the duplicated momentum components are in the
 $x$- and $y$-directions.
This could have either $p > q$ or $q > p$, depending on the momentum.

\subsection{Little Group Basis Choice}
\label{sec:lgbasis}

The first set of basis vectors that belong to the rotation little group are determined
 by applying the eigenvector equation, Eq.~(\ref{eq:irrepeigenvector}).
Different basis vectors could be deduced from different choices when
 solving the eigenvalue equation for a different choice of $R$.
Some of the basis vectors are computed from an eigenvector by application
 of representation matrices.
The choice of $R$ is different for each momentum class and little group representation.
To make this explicit, the conventional choices are outlined here.
In cases where the little group irrep is 1-dimensional,
 only one basis vector comes from the eigenvector equation
 and so no conventional choices are needed.

\begin{itemize}
\item $(0,0,0)$ $E^\pm$:\\
 $R=R_z$, with $\lambda = +1$ for $\ket{v_0}$ and $\lambda = -1$ for $\ket{v_1}$.\\
 Phase is fixed such that $\bra{v_0} R_x \ket{v_1} = -\frac{\sqrt{3}}{2}$.
\item $(0,0,0)$ $T^\pm_1$:\\
 $R=R_x$, with $\lambda = +1$ for $\ket{v_0}$.\\
 The other representation vectors are determined from the relations
 $\ket{v_1} = \lambda R_z \ket{v_0}$ and $\ket{v_2} = \lambda R_x \ket{v_1}$.
\item $(0,0,0)$ $T^\pm_2$:\\
 Same conventions as for $(0,0,0)$ $T^\pm_1$, except with $\lambda = -1$.
\item $(0,0,0)$ $G^\pm_1$:\\
 $R=R_z$, with $\lambda = \frac{1-i}{\sqrt{2}}$ for $\ket{v_0}$
 and $\lambda = \frac{1+i}{\sqrt{2}}$ for $\ket{v_1}$.\\
 Phase is fixed such that $\bra{v_0} R_x \ket{v_1} = -\frac{i}{\sqrt{2}}$.
\item $(0,0,0)$ $G^\pm_2$:\\
 $R=R_z$, with $\lambda = -\frac{1-i}{\sqrt{2}}$ for $\ket{v_0}$
 and $\lambda = -\frac{1+i}{\sqrt{2}}$ for $\ket{v_1}$.\\
 Phase is fixed such that $\bra{v_0} R_x \ket{v_1} = \frac{i}{\sqrt{2}}$.
\item $(0,0,0)$ $H^\pm$:\\
 $R=R_z$, with
 $\lambda =  \frac{1-i}{\sqrt{2}}$ for $\ket{v_0}$,
 $\lambda =  \frac{1+i}{\sqrt{2}}$ for $\ket{v_1}$,
 $\lambda = -\frac{1-i}{\sqrt{2}}$ for $\ket{v_2}$, and
 $\lambda = -\frac{1+i}{\sqrt{2}}$ for $\ket{v_3}$.\\
 Phases are fixed to
 $\bra{v_0} R_x \ket{v_1} = i\sqrt{\frac{1}{8}}$,
 $\bra{v_0} R_x \ket{v_2} = -\sqrt{\frac{3}{8}}$, and
 $\bra{v_0} R_x \ket{v_3} = i\sqrt{\frac{3}{8}}$.
\item $(0,0,p)$ $E_2$:\\
 $R=R_z$, with $\lambda = +i$ for $\ket{v_0}$.\\
 Other vector determined from $\ket{v_1} = (R_y)^2 P \ket{v_0}$.
\item $(0,0,p)$ $G_1$:\\
 $R=R_z$, with $\lambda = \frac{1-i}{\sqrt{2}}$ for $\ket{v_0}$
 and $\lambda = \frac{1+i}{\sqrt{2}}$ for $\ket{v_1}$.\\
 Phase is fixed such that $\bra{v_0} (R_y)^2 P \ket{v_1} = 1$.
\item $(0,0,p)$ $G_2$:\\
 Same as $(0,0,p)$ $G^\pm_2$ with $\lambda \to -\lambda$.
\item $(p,p,0)$ $G_1$:\\
 $R=(R_z)^2 P$, with $\lambda = +i$ for $\ket{v_0}$
 and $\lambda = -i$ for $\ket{v_1}$.\\
 Phase is fixed such that $\bra{v_0} (R_z)^{-1} (R_y)^2 \ket{v_1} = \frac{1-i}{\sqrt{2}}$.
\item $(p,p,p)$ $B$:\\
 $R=R_y R_z$, with $\lambda = \frac{-1+\sqrt{3}i}{2}$ for $\ket{v_0}$
 and $\lambda = \frac{-1-\sqrt{3}i}{2}$ for $\ket{v_1}$.\\
 Phase is fixed such that $\bra{v_0} R_z (R_y)^2 P \ket{v_1} = 1$. 
\item $(p,p,p)$ $G_1$:\\
 $R=R_z (R_y)^2 P$, with $\lambda = +i$ for $\ket{v_0}$
 and $\lambda = -i$ for $\ket{v_1}$.\\
 Phase is fixed such that $\bra{v_0} R_y R_z \ket{v_1} = (1-i)\sqrt{\frac{3}{8}}$.
\end{itemize}
All of the other little group irreps are 1-dimensional.

\subsection{Momentum Class Cosets}
\label{sec:cosets}

Identifying the basis elements by applying the group theory projection to the little
 group gives a set of vectors that all transform with the reference momentum.
This is only one of the momenta that belong to any single representation.
Basis vectors for other momenta are determined by rotating the basis vector
 with representation elements that do not belong to the little group.
The representative set of elements needed to fill out every momentum choice
 is called the coset.
In mathematical language, if $G$ is the full group of representation elements
 and $K$ is the little group, then the product of elements in the coset $H$
 with all elements of $K$ should reproduce every element of $G$.
In other words,
\begin{equation}
 G = HK = \{ g = hk : \forall h \in H, k \in K \} \,.
\end{equation}

Assuming that the basis vectors $\ket{v_i(\vec{p}_{ref})}$ for $i\in \{1,\dots,\dim{\sigma}\}$
 are known from projection of the little group and the basis conventions described above,
 then the basis vectors with $\vec{p} \neq \vec{p}_{ref}$ are computed
 by applying a coset element to the vector, i.e.
\begin{equation}
 \ket{v_i(\vec{p})} = R \ket{v_i(\vec{p}_{ref})} : R \in H, \, \vec{p} = R \vec{p}_{ref} \,.
\end{equation}
Here, $H$ denotes the coset for the appropriate momentum class, which are listed below.
Computing the remaining basis vectors by requiring this relation conventionally
 fixes the basis for all irreps.
The only task that remains is to pick the identifying cosets for each momentum class,
 thereby fixing the transformations from the little group irrep to the full irrep.
A different choice of coset elements would result in a different basis.
The conventional choice of coset elements are listed below,
 in the order that they are applied in the code.

\begin{itemize}
\item $(0,0,0)$: $\{\mathds{1}\}$
\item $(0,0,p)$: $\{ R_x^{2}, R_x, R_y^{-1}, R_y, R_x^{-1}, \mathds{1} \}$
\item $(p,p,0)$: $\{
R_z^{2}, R_yR_x^{-1}, R_xR_z, R_z, R_z^{-1}R_x^{-1},
R_xR_y^{-1}, R_y, R_y^{-1}, R_z^{-1}, R_x^{-1},
R_x, \mathds{1}
\}$
\item $(p,p,p)$: $\{
R_xR_zR_y^{-1}, R_xR_z, R_yR_x^{-1},
R_z, R_xR_z^{-1}, R_x, R_y, \mathds{1}
\}$
\item $(p,q,0)$: $\{
R_z^{2}, R_xR_y^{2}, R_z^{2}R_x, R_y^{2}, R_xR_zR_y^{-1},
R_yR_x^{-1}, R_xR_z, R_z, R_z^{-1}R_x^{-1}, R_xR_y^{-1},
R_xR_z^{-1}R_x,\\
\phantom{\quad} R_zR_xR_z, R_y, R_y^{-1}, R_zR_x^{-1},
R_xR_y, R_z^{-1}, R_xR_z^{-1}, R_y^{-1}R_x^{-1}, R_yR_zR_x,
R_x^{2}, R_x^{-1}, R_x, \mathds{1}
\}$
\item $(p,p,q)$: $\{
R_xR_zR_y^{-1}, R_x^{2}, R_y^{2}, R_yR_zR_x, R_xR_y^{2},
R_xR_z^{-1}, R_xR_z^{-1}R_x, R_z^{-1}R_x^{-1}, R_zR_x^{-1}, R_y,
R_yR_x^{-1}, R_x^{-1},\\
\phantom{\quad} R_xR_z, R_x, R_xR_y^{-1},
R_zR_xR_z, R_y^{-1}, R_xR_y, R_z^{2}R_x, R_y^{-1}R_x^{-1},
R_z^{2}, R_z^{-1}, R_z, \mathds{1}
\}$
\item $(p,q,r)$: $\{
R_xR_z^{-1}R_x, R_y^{-1}P, R_y^{-1}R_x^{-1}P, R_xR_z^{-1}, R_yR_x^{-1}, R_xR_zP,
R_zR_xR_zP, R_y, R_xR_yP, R_z^{-1}R_x^{-1}, R_xR_y^{2}, R_z^{2}R_xP,\\
\phantom{\quad} R_xP, R_x^{-1}, R_zR_x^{-1}, R_xR_y^{-1}P, R_xR_zR_y^{-1}, R_zP,
P, R_x^{2}, R_y^{2}, R_z^{2}P, R_z^{-1}P, R_yR_zR_x,\\
\phantom{\quad} R_yR_zR_xP, R_z^{-1}, R_z^{2}, R_y^{2}P, R_x^{2}P, \mathds{1},
R_z, R_xR_zR_y^{-1}P, R_xR_y^{-1}, R_zR_x^{-1}P, R_x^{-1}P, R_x,\\
\phantom{\quad} R_z^{2}R_x, R_xR_y^{2}P , R_z^{-1}R_x^{-1}P, R_xR_y, R_yP, R_zR_xR_z,
R_xR_z, R_yR_x^{-1}P, R_xR_z^{-1}P, R_y^{-1}R_x^{-1}, R_y^{-1}, R_xR_z^{-1}R_xP
\}$
\end{itemize}

\bibliographystyle{apsrev4-1}
\bibliography{wickop}

\end{document}